\documentclass[aps,prl,showpacs,twocolumn,groupedaddress]{revtex4}
\usepackage[english]{babel}
\usepackage{graphicx}
\usepackage{color}
\usepackage[utf8]{inputenc}
\usepackage{pstricks,pst-grad,color}
\usepackage{graphicx,pifont,amssymb}
\usepackage{amsmath,amssymb}

\begin{document}

%\preprint{}

\title{Spin-selective Kondo insulator: Cooperation of ferromagnetism and Kondo effect} 



\author{Robert Peters}
\email[]{peters@scphys.kyoto-u.ac.jp}
\affiliation{Department of Physics, Kyoto University, Kyoto 606-8502, Japan}
\author{Norio Kawakami}
\affiliation{Department of Physics, Kyoto University, Kyoto 606-8502, Japan}
\author{Thomas Pruschke}
\affiliation{Department of Physics, University of G\"ottingen, 37077 Göttingen, Germany}

\date{\today}

%%%%%%%%%%%%%%%%%%%%%%%%%%%%%%%%%%%%%%%%%%%%%%%%%%%%%%%5
\begin{abstract}
We propose the notion of spin-selective Kondo insulator, which
provides a fundamental mechanism to describe the ferromagnetic phase
of the Kondo lattice model with antiferromagnetic coupling. This
unveils a remarkable feature of the ferromagnetic metallic phase: the
majority-spin conduction electrons show metallic- while the
minority-spin electrons show insulating-behavior. The resulting Kondo
gap in the minority spin sector, which is due to the cooperation of
ferromagnetism and partial Kondo screening, evidences a
dynamically-induced commensurability for a combination of
minority-spin electrons and parts of localized spins. Furthermore,
this mechanism predicts a nontrivial relation between the macroscopic
quantities such as electron magnetization, spin polarization and
electron filling. 
\end{abstract}
%%%%%%%%%%%%%%%%%%%%%%%%%%%%%%%%%%%%%%%%%%%%%%%%%%%%%%%%%%%%

%\keywords{}

\maketitle

Even 30 years after their discovery heavy-fermion systems attract much
attention due to their fascinating properties. Apart from being Fermi
liquids with effective mass thousand times as large as
 the free electron one,
they show all kinds of competing or coexisting phases, and at the
boundaries between these phases one frequently observes quantum phase
transitions, accompanied by barely understood non-Fermi liquid
behavior \cite{coleman2007,coleman2005,gegenwart2008}.
Heavy fermion compounds usually include lanthanides or
actinides with open 4$f$- or 5$f$-shells, which in the simplest theoretical
modeling can be viewed as a regular lattice of local moments coupled
to the conduction electrons. This coupling typically leads to two 
competing mechanisms: the long-ranged RKKY interaction
and  the local Kondo screening.  While the RKKY
interaction favors a 
magnetically ordered state, the Kondo screening is usually considered to form a
paramagnetic heavy-fermion state. The competition of these two mechanisms
can be easily understood in terms of the Doniach phase diagram \cite{doniach77}.

While in most heavy-fermion compounds the magnetic order is
antiferromagnetic, there are a certain class of compounds showing
ferromagnetic order. 
For example, the recently discovered YbNi$_4$P$_2$ 
is a ferromagnetically ordered heavy-fermion compound which seems to
be very close to a quantum critical point \cite{krellner2011}.
Taking such ferromagnetic heavy fermion compounds as
  motivation we analyze in detail the mechanism stabilizing the
  ferromagnetic state. 
An interesting question in this context is, if and how the Kondo
effect accounts for the ferromagnetic
state \cite{lee1988,perkins2007,yamamoto2010,li2010}.

In this letter, we propose a {\it spin-selective Kondo insulator}, where 
the Kondo screening plays an essential role in stabilizing the ferromagnetic 
metallic state at zero temperature, which elucidates a 
previously unrecognized  feature of the ferromagnetic phase: the majority-spin (minority-spin) conduction electrons are in a metallic (insulating) state. We claim that this notion is not 
specific to certain choices of system parameters but is 
fundamental and ubiquitous for the ferromagnetic phase in 
the Kondo lattice model.  Due to partial Kondo screening, parts of 
the local moments are bound to the electrons, resulting in a 
dynamically-induced commensurability which is essential for 
producing the gap in the minority spin electrons. We find that this 
commensurability condition leads to a nontrivial relation between 
electron magnetization, spin polarization and electron filling.

The competition or cooperation between the magnetic phase mediated by
the RKKY interaction and
the Kondo screening can be modeled via a Kondo lattice model with
antiferromagnetic coupling between the local moments and the conduction
 electrons. 
The Kondo lattice model reads \cite{doniach77,lacroix1979,fazekas1991},
\begin{eqnarray}
H&=&t\sum_{<i,j>\sigma}c^\dagger_{i\sigma}c_{j\sigma}+J\sum_i\vec{S}_i\vec{s}_i\nonumber\\
\vec{s}_i&=&c^\dagger_{i\sigma_m}\vec{\rho}_{\sigma_m\sigma_n}c_{i\sigma_n}\nonumber,
\end{eqnarray}
where $c^\dagger_{i\sigma}$ creates an electron on site $i$ with
spin-direction $\sigma$, $\vec{\rho}$ represents the vector of Pauli-matrices, 
and $\vec{S}_i$ represents the local spins
which are coupled to the electrons via an antiferromagnetic spin-spin
interaction with strength $J>0$. 

To solve the Kondo lattice model we use the dynamical mean field
theory (DMFT) \cite{metzner1989,georges1996,pruschke1995}. DMFT maps
the lattice model onto a quantum impurity 
model with a fermionic bath being determined self-consistently. 
Although being an approximation 
to real systems, DMFT has provided many insights into the 
physical properties and can even captures subtle differences in the
lattice geometry.
For solving the impurity model, we use the numerical
renormalization group (NRG) \cite{wilson1975,bulla2008}, which is able
to reliably calculate spectral 
functions at very low temperatures \cite{peters2006,weichselbaum2007}.

First, we briefly summarize the known DMFT results for the
Kondo lattice model 
\cite{lacroix1979,fazekas1991,peters2007,otsuki2009} (A
  discussion on the RKKY interaction within DMFT can be found in
  \cite{peters2007}.) 
At half filling there is
a pronounced antiferromagnetic N\'eel state for weak coupling, which
vanishes with increasing coupling strength $J$ via a
continuous transition to a paramagnetic insulating state,
 the Kondo insulator.
Doping slightly away from half filling this transition changes
into a transition between an antiferromagnetic state (possibly
spin-density-wave) and a paramagnetic metallic state.
Especially the paramagnetic state
around half filling is dominated by the Kondo effect, where 
the Kondo screening of localized spins results in a large Fermi
surface accompanied by a narrow band and a gap close to the Fermi energy.
Away from half filling the effects of Kondo screening become 
less important as there is an imbalance
between local moments and available conduction electrons. 
Such a tendency might be even stronger
when the system enters a ferromagnetic state realized at
low fillings, because
an additional imbalance between spin-up and spin-down
electrons arises. Contrary to this naive expectation, however,
we demonstrate here that the Kondo screening plays an essential 
role even in the ferromagnetic phase. In particular,
we reveal that the cooperation of ferromagnetism 
and Kondo screening can realize a novel kind of Kondo 
insulating state in the ferromagnetic metallic phase.

\begin{figure}[tb]
\includegraphics[width=1\linewidth]{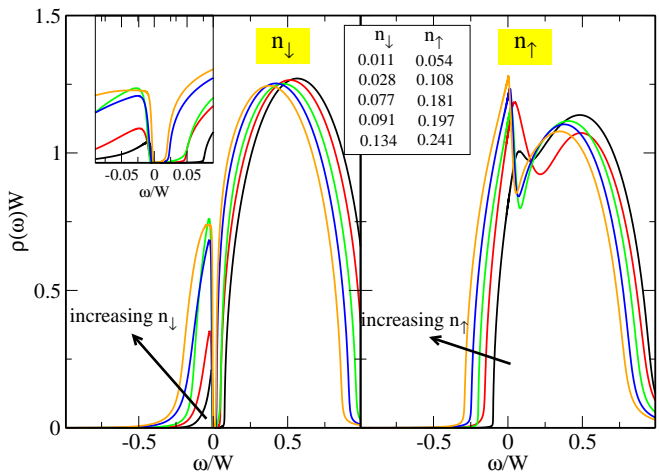}
\caption{(Color online) Spin-resolved spectral functions for the
  ferromagnetic state in the Kondo lattice model
  $J/W=0.25$. The inset shows a magnification around the Fermi energy
  for the spin-down component illustrating the gap in the spectral function.
\label{fig1}}
\end{figure}
Figure \ref{fig1} shows the local spin-resolved spectral-functions calculated 
in the ferromagnetic phase for a Bethe lattice with antiferromagnetic 
Kondo coupling $J/W=0.25$ (bandwidth $W=4t$). For this coupling strength
the ferromagnetic phase extends from a nearly empty system to
approximately $n^c=n^c_\uparrow+n^c_\downarrow=0.5$. 
One finds a striking difference in the spectral functions, which has
not been recognized previously, for the majority spin ($n^c_\uparrow$) 
and the minority spin ($n^c_\downarrow$). While in the majority-spin spectral
function a peak at
the Fermi energy $\omega=0$ and a dip for $\omega>0$ can be found,
{\it there is a gap at the Fermi energy in the minority-spin spectral
function}. It is important to note that such a gap is not present in
the ferromagnetic phase for a Kondo lattice model with
ferromagnetically coupled spins. We propose that this gap in the spectral 
function is due to a partial Kondo screening of the localized spins,
which results in an intriguing state: although the ferromagnetic state is 
metallic, only the majority-spin electrons contribute to the low-temperature
properties, in particular transport. The minority spins, even though
not completely depleted, form an insulator, which we name {\it spin-selective
Kondo insulator}. 

Increasing the occupation number, the dip in the
majority-spin spectral function moves closer to the Fermi energy and
becomes more pronounced. Eventually, the ferromagnetic state is replaced 
by a paramagnetic state, for which the spectral functions for
 both spin-components suffer from 
the typical suppression of the DOS for $\omega>0$ due to Kondo screening.
Increasing the occupation towards half filling this
dip becomes deeper and finally moves to the Fermi energy,
forming the Kondo insulator.

\begin{figure}[t]
\includegraphics[width=1\linewidth]{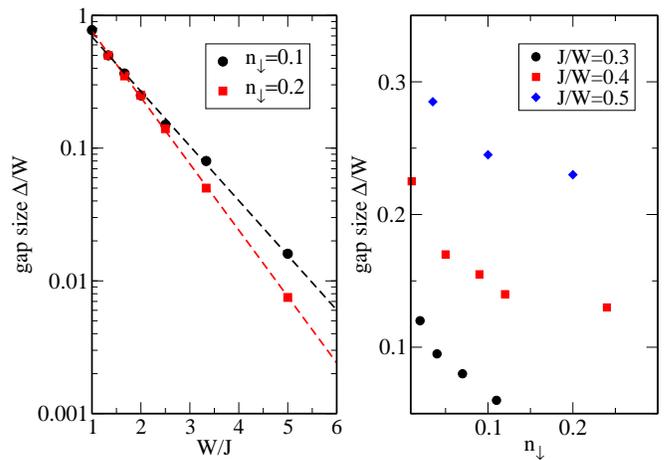}
\caption{(Color online) Gap width $\Delta/W$ in the minority-spin
  spectral function depending
  on the spin-coupling $J$ and the occupation $n^c_\downarrow$. The
  temperature of the system is $T/W=3\cdot 10^{-4}$. The
  lines in the left panel are fits as $\sim\exp(-a/J)$. \label{fig2}}
\end{figure}

Clear evidence showing that the above insulating gap is indeed
caused by the Kondo screening can be found in the dependence of the gap width
$\Delta$ on the occupation and coupling strength, shown in Fig. \ref{fig2}. 
The left panel in
Fig. \ref{fig2} displays the dependence of the gap width on the coupling
strength. For this purpose the minority-spin occupancy was kept
constant (also resulting in a nearly constant majority-spin occupancy).
The dependency on $J$ perfectly obeys a Kondo temperature-like form
$\Delta\sim\exp(-a/J)$ with a fitting constant $a$, suggesting
that the Kondo physics is essential for the gap-formation.
As a function of increasing filling the gap width decreases monotonically, as
shown in the right panel of Fig. \ref{fig2}. As soon as the
ferromagnetic phase vanishes,  
the gap at the Fermi energy closes, too, and the minority and majority
spectral functions look similar to the right panel in Fig. \ref{fig1}.

\begin{figure}[tb]
\includegraphics[width=1\linewidth]{fig3a.eps}
\center{\includegraphics[width=0.6\linewidth]{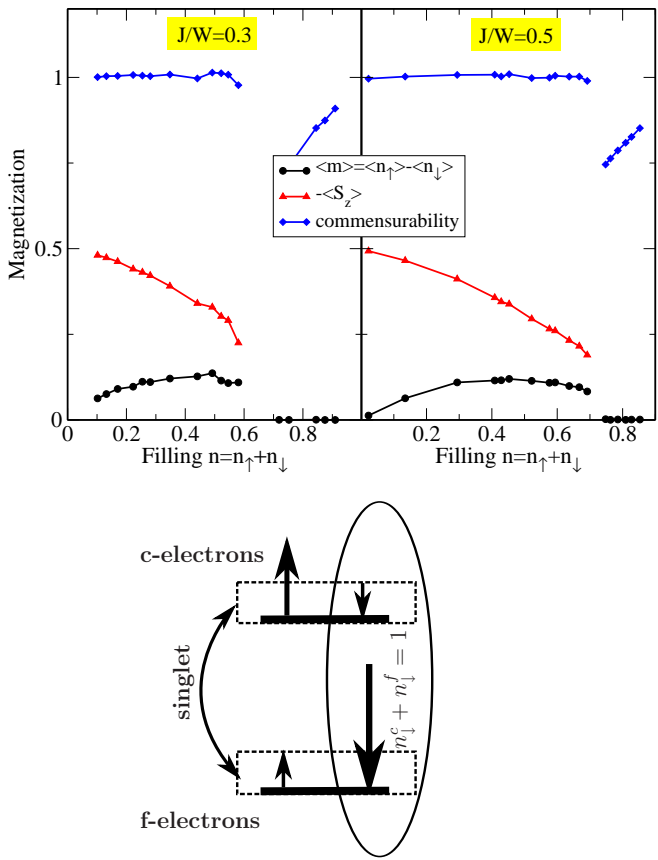}}
\caption{(Color online) Upper panel: Magnetization 
and ``commensurability'', ($n^c_\downarrow+n^f_{\downarrow}$),
for two different
  coupling strengths and different occupation numbers calculated for a
  Bethe lattice at $T/W=3\cdot 10^{-4}$. The electron
  magnetization is shown as $\langle m\rangle=n^c_\uparrow-n^c_\downarrow$, while the
  spin expectation value $\langle S_z\rangle$ is shown mirrored as
  $-\langle Sz\rangle$. The commensurability-condition is explained in
the text. Lower panel: sketch of the local configuration 
(see text).\label{fig3}}
\end{figure}
Let us now elucidate the basic physics behind this ferromagnetic
state. In Fig. \ref{fig3} the magnetization of the conduction
electrons 
$\langle m\rangle=n^c_\uparrow-n^c_\downarrow$ and the polarization of
the localized spins $-\langle S_z\rangle$ is shown. Note that
the local spin-polarization always has the sign opposite to the
conduction electron
magnetization due to the antiferromagnetic coupling, thus
$-\langle S_z\rangle$ has the same sign. Increasing the number of
conduction electrons, the 
magnetization of the electrons first increases due to increasing
filling, and eventually decreases again due to the suppression of
the ferromagnetic state. On the other hand, the spins are
almost fully polarized for a nearly empty lattice, with
monotonically decreasing polarization for increasing conduction
electron number. In the spirit of a pseudo-fermion representation, let
us assume that the localized spins are 
actually formed by a local half-filled 
and strongly-interacting energy level so that $\langle S_z\rangle=
(n^f_{\uparrow}-n^f_{\downarrow})/2$ and
$n^f_{\uparrow}+n^f_{\downarrow}=1$ (defining $n^f_{\sigma}$ as the
spin-dependent occupation of this level). 
Remarkably, we find that the following nontrivial {\it 
commensurability condition} holds within the ferromagnetic state:
\begin{equation}
n^c_\downarrow+n^f_{\downarrow}=1\label{eqsum},
\end{equation}
as can be seen in Fig. \ref{fig3}. Note that Eq. (\ref{eqsum}) is
equivalent to $n^f_{\uparrow}=n^c_\downarrow$.
It should be noticed that this condition is not {\it a priori} given but 
is generated dynamically due to many-body effects. 
To clarify the origin of the above commensurability
 we propose that a partial local Kondo-singlet is formed
in which $\langle n^c_\downarrow\rangle$ majority- and
minority-electrons participate, thus combining all spin-down
conduction electrons together with a part of 
the $f$-electrons and the spin-up conduction electrons
to a Kondo spin-singlet. Here, we have assumed that spin-down is the
minority-spin direction. The remaining 
majority-spin conduction electrons and spin-down $f$-electrons form a
ferromagnetic state. (see a sketch in the lower panel of Fig. \ref{fig3}).
That the number of spin-down electrons including $f$- and
conduction-electrons sums up to unity gives a commensurable situation, 
which results in a gap at the Fermi
energy. On the other hand, for the majority-spins there is not such a
commensurability condition, but 
$n^c_\uparrow+n^f_\uparrow=n^c=n^c_\uparrow+n^c_\downarrow$ holds.
Therefore, this partial Kondo screening results in an insulating state for the
minority-spin, while the majority-spin electrons remain metallic.
For this reason we have called this state a ``spin-selective Kondo insulator''. 
The commensurability condition (\ref{eqsum}) 
smoothly connects to the Kondo
insulator at half filling, suggesting that this ferromagnetic state
should exist up to half filling. However, our results clearly show
that there is a transition from this
ferromagnetic phase to a paramagnetic state at electron fillings
for $n^c>n_\text{ferro}$. This is only possible, if the
expectation value $\langle Sz\rangle$ jumps, leading to a
discontinuous 
phase transition at $n_\text{ferro}$ .
Our finding of the discontinuous transition completely
  agrees with the 
recent analytical results showing that non-analytic terms prevent the
continuous transition from a ferromagnet to a paramagnet
\cite{efremov2008}.

A further important consequence deduced directly from the
commensurability condition (\ref{eqsum})  is a nontrivial 
relation between
electron magnetization, spin polarization and occupation number:
\begin{equation}
2\langle Sz\rangle+\langle m\rangle=\langle n^c\rangle -1.
\label{macro}
\end{equation}
This formula
connects these three quantities which are otherwise independent from
each other. By arranging itself in this way the system can gain an
additional energy originating from the partial Kondo screening. Note
that it should be possible to verify such a relation experimentally.  

\begin{figure}[tb]
\includegraphics[width=0.49\linewidth]{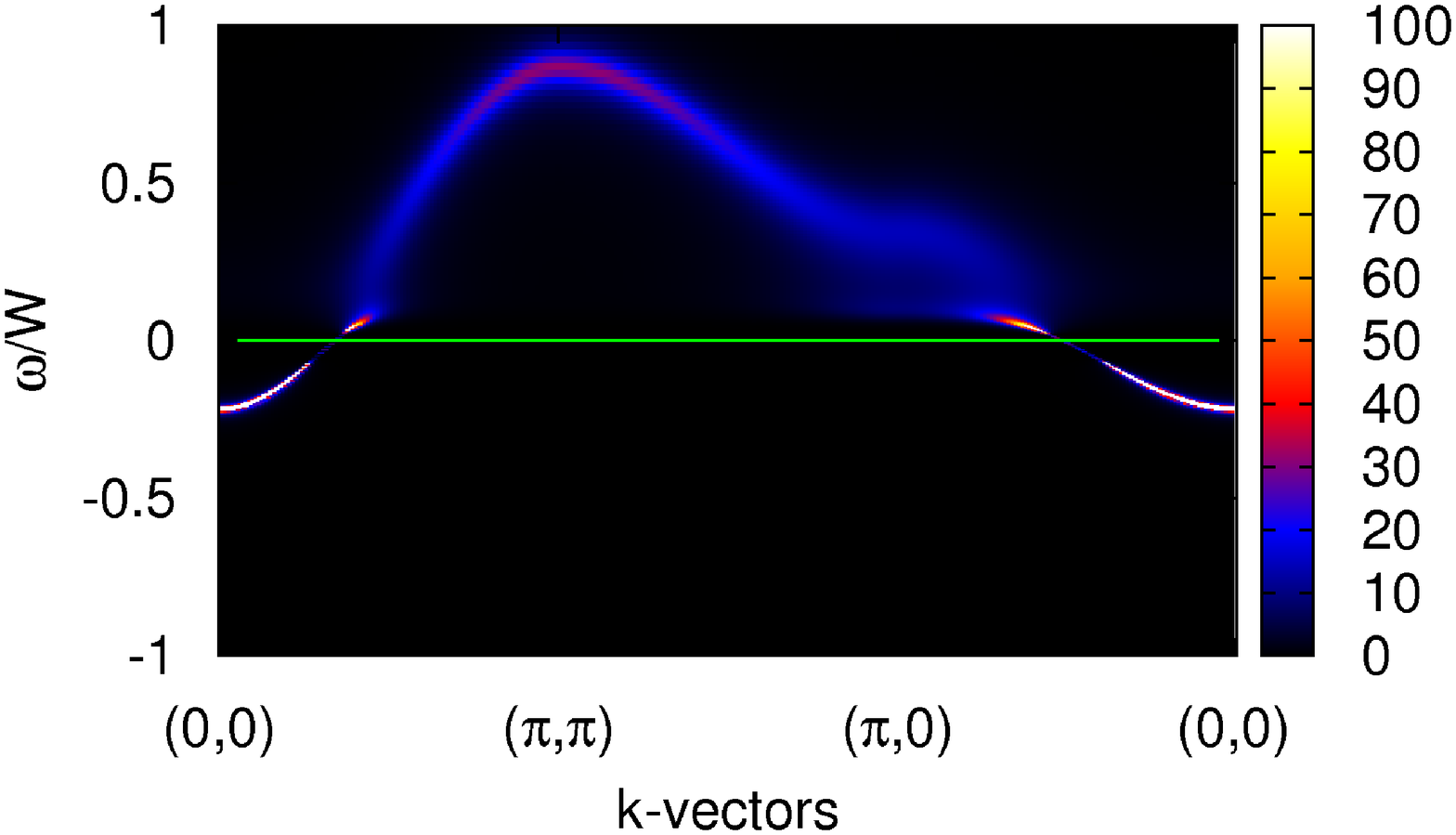}
\includegraphics[width=0.49\linewidth]{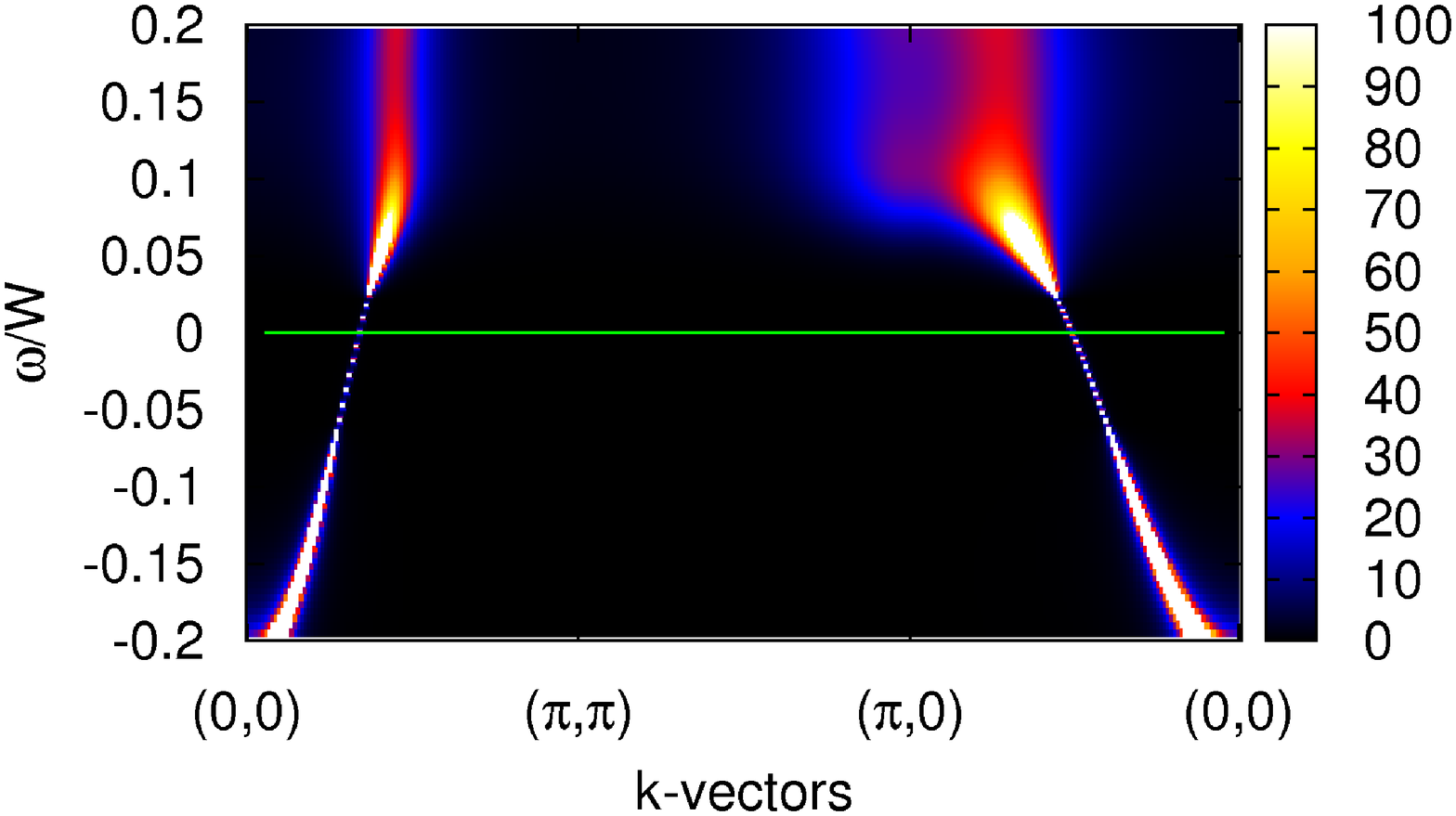}
\includegraphics[width=0.49\linewidth]{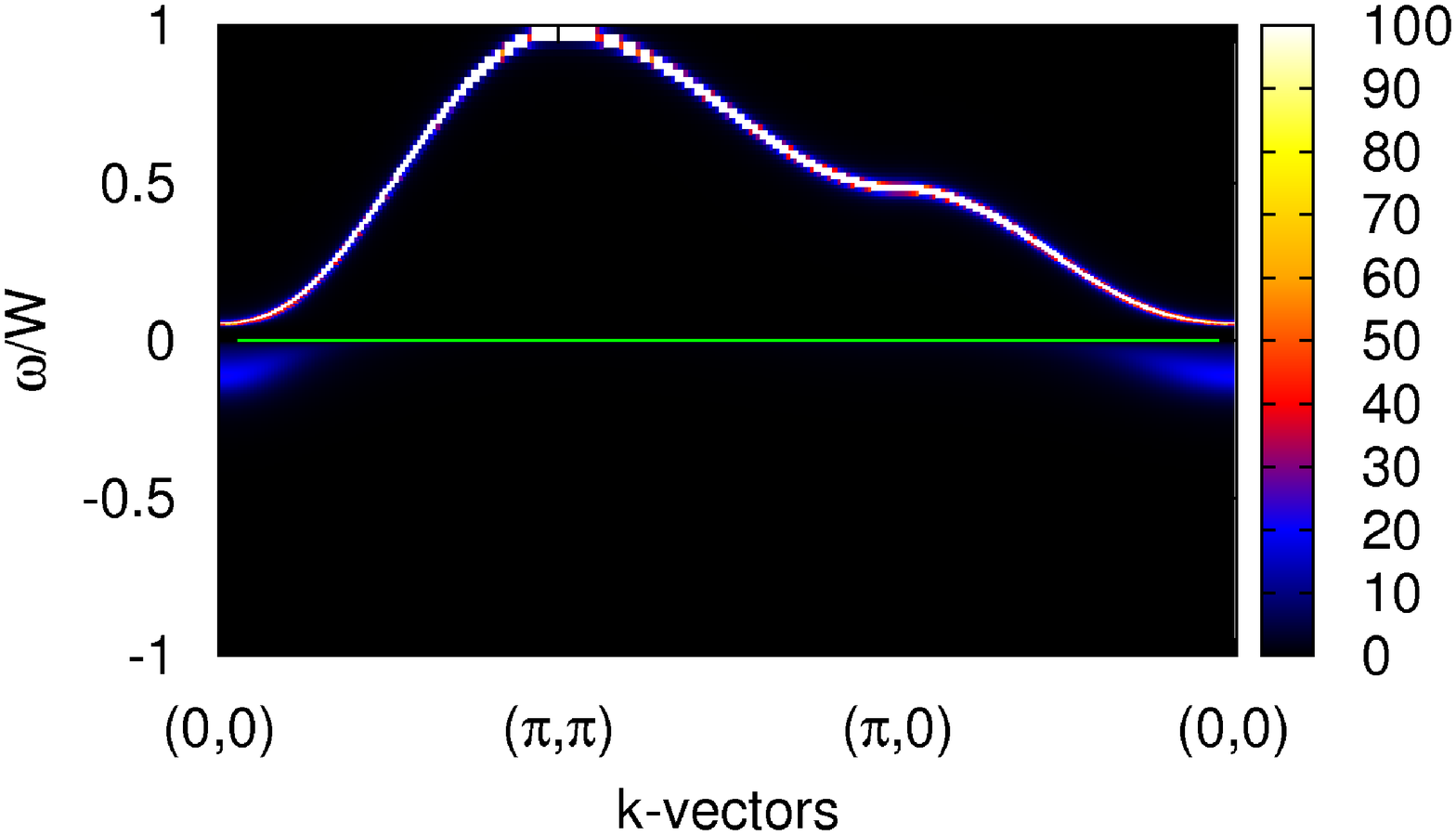}
\includegraphics[width=0.49\linewidth]{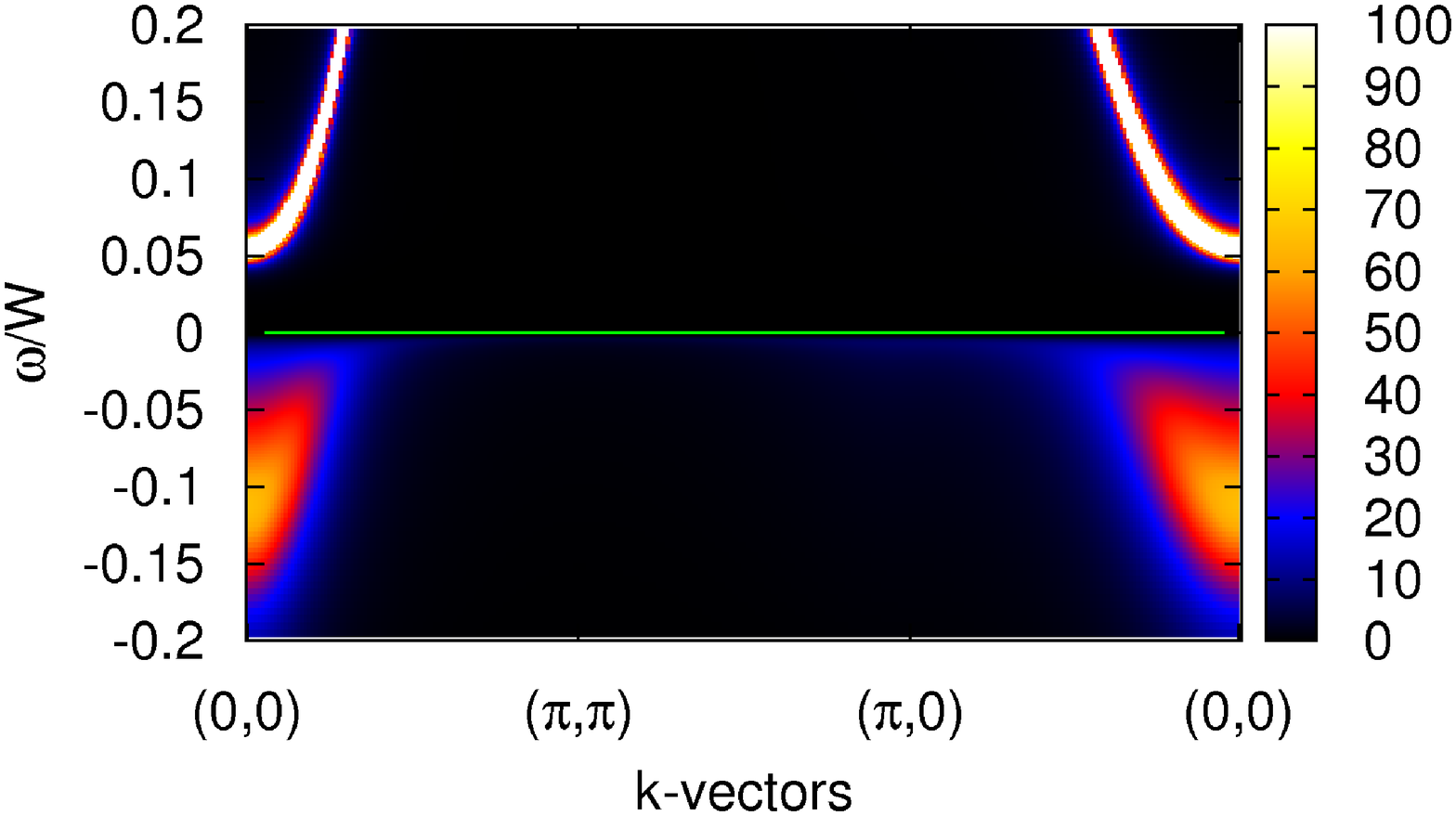}
\caption{(Color online) Momentum-resolved spectral functions for the square
  lattice and $J/W=0.3$, $n=n^c_\uparrow+n^c_\downarrow=0.25$, $m=n^c_\uparrow-n^c_\downarrow=0.1$. The right side always shows a
  magnification of the left side around the Fermi energy
  $\omega=0$ represented by the green line. From top to bottom the
  figures show the majority-spin and the minority-spin spectral
  function, respectively.
   \label{fig4}}
\end{figure}
\begin{figure}[tb]
\includegraphics[width=0.49\linewidth]{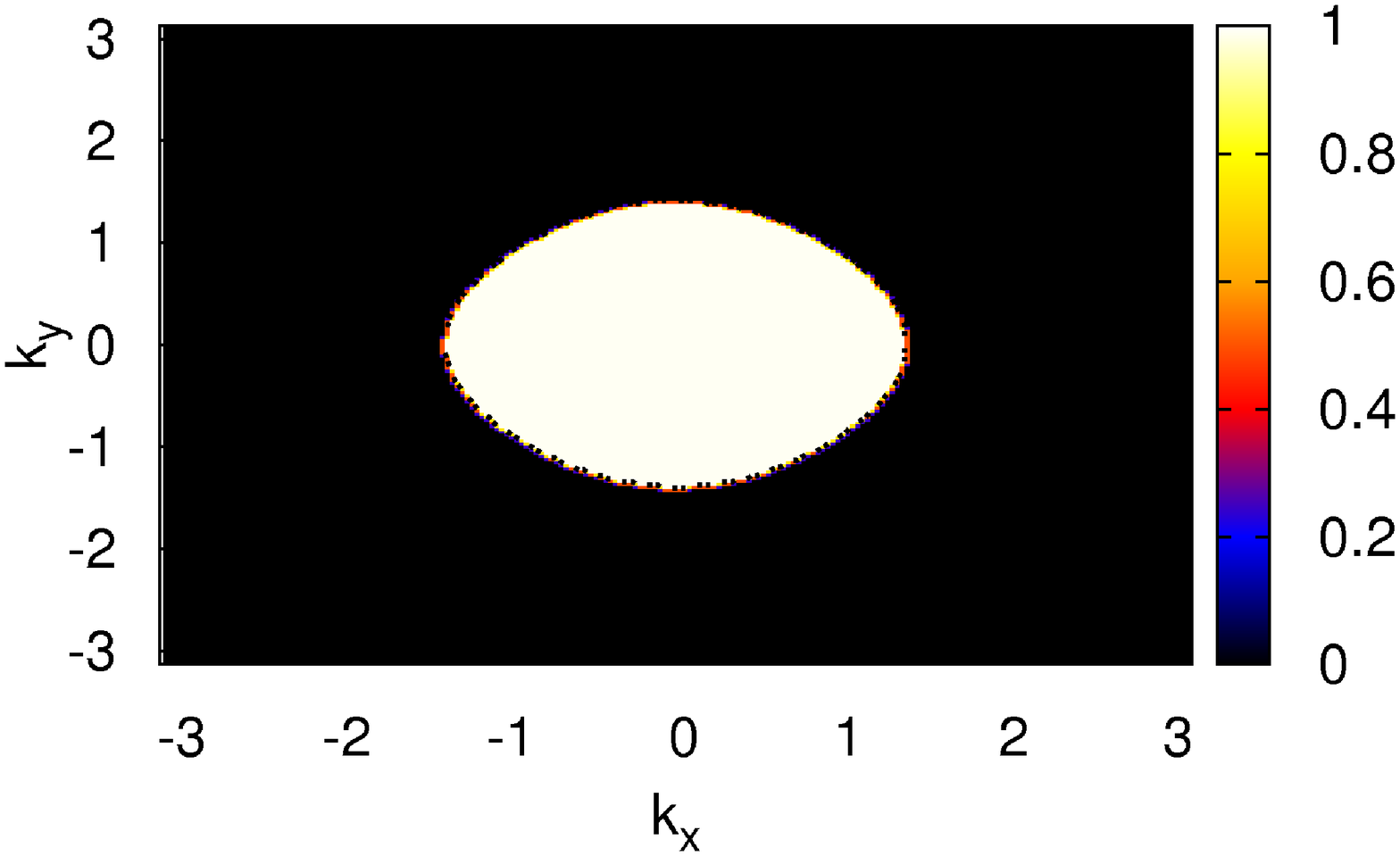}
\includegraphics[width=0.49\linewidth]{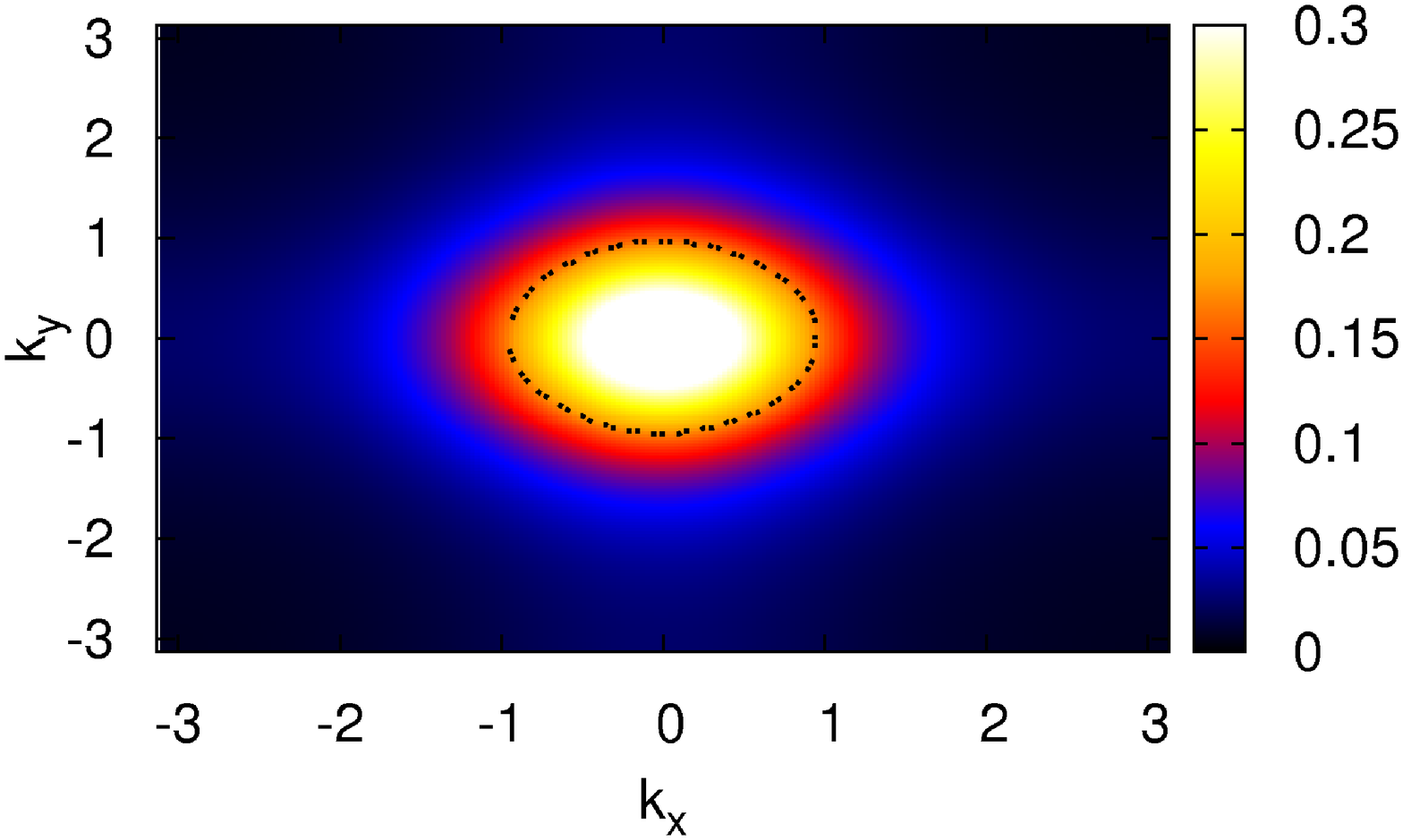}
\caption{(Color online) Momentum-resolved occupation number $n(k)$ (same 
  parameters as in Fig. \ref{fig4}). Left (right) panel shows the
  majority- (minority-) spin component. The dotted line represents the
  Fermi surface for non-interacting electrons (for the majority-spin
  this lines coincides with the shown surface). Note that for improving
  the contrast, the occupation for the minority-spin electrons is
  displayed in the interval $n(k)\in[0,0.3]$.
 \label{fig5}}
\end{figure}
The formation of the gap in the spectral function does not depend on
the lattice geometry. For example, it can also be found in
DMFT calculations for a two-dimensional square lattice. Figure
\ref{fig4} shows the momentum-resolved spectral functions for $J/W=0.3$ and
$n^c_\uparrow+n^c_\downarrow=0.25$ for a square lattice. The right panels
are magnifications around the Fermi energy. In the
minority-spin spectral 
function (bottom right) the gap can be clearly seen. 
While the majority-spin electrons are renormalized for $\omega>0$
with a finite life-time, the minority-spin electrons are
renormalized for $\omega<0$. This behavior is also visualized in
Fig. \ref{fig5}, in which the momentum-resolved occupation number is
shown. While for the majority-spin electrons the occupation number
distribution looks like in the non-interacting case, for the
minority-spin electrons this function
is actually smeared out as compared to a reference non-interacting
system, i.e. we indeed observe a large Fermi volume here.

Using DMRG, we have confirmed that the spin-selective
  Kondo insulator 
together with the commensurability condition can be found for the
ferromagnetic phase of the one-dimensional (1D) Kondo lattice model, too.
In fact, previous calculations for the ferromagnetic ground state
observed a magnetization $S_{tot}=1/2(L-N_c)$ \cite{tsunetsugu1997}
($L$ system length, $N_c$ electron number), which supports the
commensurability condition, and two separated bands in
the spectral functions \cite{smerat2009}. From these precise analyses
of the 1D model, we conclude that our finding is ubiquitous for the
Kondo lattice model and not an artifact of DMFT.

In conclusion we have clarified the physics behind  
 the ferromagnetic metallic phase realized in the Kondo lattice model.
We have demonstrated that the cooperation of ferromagnetism and partial 
Kondo screening results in an intriguing phase, here named  
spin-selective Kondo insulator, where an insulating state
is stabilized  for the minority-spin electrons while  
 the majority-spin electrons are still metallic. We believe that 
 the mechanism proposed here, the dynamically generated commensurability, 
should be generic for the ferromagnetic phase in
the Kondo lattice models. It alternatively provides 
the nontrivial relation between the electron magnetization,
spin polarization and occupation number, for which the system can gain
a maximum of additional energy. The proposed relation 
between the macroscopic quantities might be confirmed in experiments.
Good candidates in this context are ferromagnetic heavy fermion
compounds, especially 
compounds having a large Kondo temperature. 
For such compounds a verification of the above stated relation might
be possible. Furthermore,
spin-resolved transport measurements should show  metallic
majority-spin but insulating minority-spin electrons as well as a
large Fermi surface for the minority-spin component.

%\begin{acknowledgments}
We acknowledge fruitful discussions with A. Koga.
RP thanks the Japan Society for the Promotion of Science (JSPS)
and the Alexander von Humboldt-Foundation. TP also gratefully
acknowledges support by JSPS through the Bridge program. NK is
supported by KAKENHI 
(Nos. 21540359, 20102008) and JSPS through its FIRST Program.
%\end{acknowledgments}

%\bibliography{paper}

\begin{thebibliography}{23}
\expandafter\ifx\csname natexlab\endcsname\relax\def\natexlab#1{#1}\fi
\expandafter\ifx\csname bibnamefont\endcsname\relax
  \def\bibnamefont#1{#1}\fi
\expandafter\ifx\csname bibfnamefont\endcsname\relax
  \def\bibfnamefont#1{#1}\fi
\expandafter\ifx\csname citenamefont\endcsname\relax
  \def\citenamefont#1{#1}\fi
\expandafter\ifx\csname url\endcsname\relax
  \def\url#1{\texttt{#1}}\fi
\expandafter\ifx\csname urlprefix\endcsname\relax\def\urlprefix{URL }\fi
\providecommand{\bibinfo}[2]{#2}
\providecommand{\eprint}[2][]{\url{#2}}

\bibitem[{\citenamefont{Coleman}(2007)}]{coleman2007}
\bibinfo{author}{\bibfnamefont{P.}~\bibnamefont{Coleman}},
  \emph{\bibinfo{title}{Handbook of Magnetism and Advanced Magnetic Materials}}
  (\bibinfo{publisher}{John Wiley and Sons}, \bibinfo{year}{2007}),
  p.~\bibinfo{pages}{95}.

\bibitem[{\citenamefont{Coleman and Schofield}(2005)}]{coleman2005}
\bibinfo{author}{\bibfnamefont{P.}~\bibnamefont{Coleman}} \bibnamefont{and}
  \bibinfo{author}{\bibfnamefont{A.}~\bibnamefont{Schofield}},
  \bibinfo{journal}{Nature} \textbf{\bibinfo{volume}{443}},
  \bibinfo{pages}{226} (\bibinfo{year}{2005}).

\bibitem[{\citenamefont{Gegenwart and Si}(2008)}]{gegenwart2008}
\bibinfo{author}{\bibfnamefont{P.}~\bibnamefont{Gegenwart}} \bibnamefont{and}
  \bibinfo{author}{\bibfnamefont{Q.}~\bibnamefont{Si}},
  \bibinfo{journal}{Nature Physics} \textbf{\bibinfo{volume}{4}},
  \bibinfo{pages}{186} (\bibinfo{year}{2008}).

\bibitem[{\citenamefont{Doniach}(1977)}]{doniach77}
\bibinfo{author}{\bibfnamefont{S.}~\bibnamefont{Doniach}},
  \bibinfo{journal}{Physica B} \textbf{\bibinfo{volume}{91}},
  \bibinfo{pages}{231} (\bibinfo{year}{1977}).

\bibitem[{\citenamefont{Krellner et~al.}(2011)\citenamefont{Krellner, Lausberg,
  Steppke, Brando, Pedrero, Pfau, Tencé, Rosner, Steglich, and
  Geibel}}]{krellner2011}
\bibinfo{author}{\bibfnamefont{C.}~\bibnamefont{Krellner}},
  \bibinfo{author}{\bibfnamefont{S.}~\bibnamefont{Lausberg}},
  \bibinfo{author}{\bibfnamefont{A.}~\bibnamefont{Steppke}},
  \bibinfo{author}{\bibfnamefont{M.}~\bibnamefont{Brando}},
  \bibinfo{author}{\bibfnamefont{L.}~\bibnamefont{Pedrero}},
  \bibinfo{author}{\bibfnamefont{H.}~\bibnamefont{Pfau}},
  \bibinfo{author}{\bibfnamefont{S.}~\bibnamefont{Tencé}},
  \bibinfo{author}{\bibfnamefont{H.}~\bibnamefont{Rosner}},
  \bibinfo{author}{\bibfnamefont{F.}~\bibnamefont{Steglich}}, \bibnamefont{and}
  \bibinfo{author}{\bibfnamefont{C.}~\bibnamefont{Geibel}},
  \bibinfo{journal}{New Journal of Physics} \textbf{\bibinfo{volume}{13}},
  \bibinfo{pages}{103014} (\bibinfo{year}{2011}).

\bibitem[{\citenamefont{Lee et~al.}(1988)\citenamefont{Lee, Ku, and
  Shelton}}]{lee1988}
\bibinfo{author}{\bibfnamefont{W.}~\bibnamefont{Lee}},
  \bibinfo{author}{\bibfnamefont{H.}~\bibnamefont{Ku}}, \bibnamefont{and}
  \bibinfo{author}{\bibfnamefont{R.}~\bibnamefont{Shelton}},
  \bibinfo{journal}{Phys.~Rev.~B} \textbf{\bibinfo{volume}{38}},
  \bibinfo{pages}{11562} (\bibinfo{year}{1988}).

\bibitem[{\citenamefont{Perkins et~al.}(2007)\citenamefont{Perkins, Iglesias,
  Nunez-Regueiro, and Coqblin}}]{perkins2007}
\bibinfo{author}{\bibfnamefont{N.}~\bibnamefont{Perkins}},
  \bibinfo{author}{\bibfnamefont{J.}~\bibnamefont{Iglesias}},
  \bibinfo{author}{\bibfnamefont{M.}~\bibnamefont{Nunez-Regueiro}},
  \bibnamefont{and} \bibinfo{author}{\bibfnamefont{B.}~\bibnamefont{Coqblin}},
  \bibinfo{journal}{Europhysics Letters} \textbf{\bibinfo{volume}{79}},
  \bibinfo{pages}{57006} (\bibinfo{year}{2007}).

\bibitem[{\citenamefont{Yamamoto and Si}(2010)}]{yamamoto2010}
\bibinfo{author}{\bibfnamefont{S.}~\bibnamefont{Yamamoto}} \bibnamefont{and}
  \bibinfo{author}{\bibfnamefont{Q.}~\bibnamefont{Si}}, \bibinfo{journal}{Proc.
  Natl. Acad. Sci. USA} \textbf{\bibinfo{volume}{107}}, \bibinfo{pages}{15704}
  (\bibinfo{year}{2010}).

\bibitem[{\citenamefont{Li et~al.}(2010)\citenamefont{Li, Zhang, and
  Yu}}]{li2010}
\bibinfo{author}{\bibfnamefont{G.}~\bibnamefont{Li}},
  \bibinfo{author}{\bibfnamefont{G.}~\bibnamefont{Zhang}}, \bibnamefont{and}
  \bibinfo{author}{\bibfnamefont{L.}~\bibnamefont{Yu}},
  \bibinfo{journal}{Phys.~Rev.~B} \textbf{\bibinfo{volume}{81}},
  \bibinfo{pages}{094420} (\bibinfo{year}{2010}).

\bibitem[{\citenamefont{Lacroix and Cyrot}(1979)}]{lacroix1979}
\bibinfo{author}{\bibfnamefont{C.}~\bibnamefont{Lacroix}} \bibnamefont{and}
  \bibinfo{author}{\bibfnamefont{M.}~\bibnamefont{Cyrot}},
  \bibinfo{journal}{Phys.~Rev.~B} \textbf{\bibinfo{volume}{20}},
  \bibinfo{pages}{1969} (\bibinfo{year}{1979}).

\bibitem[{\citenamefont{Fazekas and Muller-Hartmann}(1991)}]{fazekas1991}
\bibinfo{author}{\bibfnamefont{P.}~\bibnamefont{Fazekas}} \bibnamefont{and}
  \bibinfo{author}{\bibfnamefont{E.}~\bibnamefont{Muller-Hartmann}},
  \bibinfo{journal}{Z. Phys. B: Condens. Matter} \textbf{\bibinfo{volume}{85}},
  \bibinfo{pages}{285} (\bibinfo{year}{1991}).

\bibitem[{\citenamefont{Metzner and Vollhardt}(1989)}]{metzner1989}
\bibinfo{author}{\bibfnamefont{W.}~\bibnamefont{Metzner}} \bibnamefont{and}
  \bibinfo{author}{\bibfnamefont{D.}~\bibnamefont{Vollhardt}},
  \bibinfo{journal}{Phys.~Rev.~Lett.} \textbf{\bibinfo{volume}{62}},
  \bibinfo{pages}{324} (\bibinfo{year}{1989}).

\bibitem[{\citenamefont{Georges et~al.}(1996)\citenamefont{Georges, Kotliar,
  Krauth, and Rozenberg}}]{georges1996}
\bibinfo{author}{\bibfnamefont{A.}~\bibnamefont{Georges}},
  \bibinfo{author}{\bibfnamefont{G.}~\bibnamefont{Kotliar}},
  \bibinfo{author}{\bibfnamefont{W.}~\bibnamefont{Krauth}}, \bibnamefont{and}
  \bibinfo{author}{\bibfnamefont{M.}~\bibnamefont{Rozenberg}},
  \bibinfo{journal}{Rev.~Mod.~Phys.} \textbf{\bibinfo{volume}{68}},
  \bibinfo{pages}{13} (\bibinfo{year}{1996}).

\bibitem[{\citenamefont{Pruschke et~al.}(1995)\citenamefont{Pruschke, Jarrell,
  and Freericks}}]{pruschke1995}
\bibinfo{author}{\bibfnamefont{T.}~\bibnamefont{Pruschke}},
  \bibinfo{author}{\bibfnamefont{M.}~\bibnamefont{Jarrell}}, \bibnamefont{and}
  \bibinfo{author}{\bibfnamefont{J.}~\bibnamefont{Freericks}},
  \bibinfo{journal}{Adv.~Phys.} \textbf{\bibinfo{volume}{44}},
  \bibinfo{pages}{187} (\bibinfo{year}{1995}).

\bibitem[{\citenamefont{Wilson}(1975)}]{wilson1975}
\bibinfo{author}{\bibfnamefont{K.}~\bibnamefont{Wilson}},
  \bibinfo{journal}{Rev.~Mod.~Phys.} \textbf{\bibinfo{volume}{47}},
  \bibinfo{pages}{773} (\bibinfo{year}{1975}).

\bibitem[{\citenamefont{Bulla et~al.}(2008)\citenamefont{Bulla, Costi, and
  Pruschke}}]{bulla2008}
\bibinfo{author}{\bibfnamefont{R.}~\bibnamefont{Bulla}},
  \bibinfo{author}{\bibfnamefont{T.}~\bibnamefont{Costi}}, \bibnamefont{and}
  \bibinfo{author}{\bibfnamefont{T.}~\bibnamefont{Pruschke}},
  \bibinfo{journal}{Rev.~Mod.~Phys.} \textbf{\bibinfo{volume}{80}},
  \bibinfo{pages}{395} (\bibinfo{year}{2008}).

\bibitem[{\citenamefont{Peters et~al.}(2006)\citenamefont{Peters, Pruschke, and
  Anders}}]{peters2006}
\bibinfo{author}{\bibfnamefont{R.}~\bibnamefont{Peters}},
  \bibinfo{author}{\bibfnamefont{T.}~\bibnamefont{Pruschke}}, \bibnamefont{and}
  \bibinfo{author}{\bibfnamefont{F.}~\bibnamefont{Anders}},
  \bibinfo{journal}{Phys.~Rev.~B} \textbf{\bibinfo{volume}{74}},
  \bibinfo{pages}{245114} (\bibinfo{year}{2006}).

\bibitem[{\citenamefont{Weichselbaum and von Delft}(2007)}]{weichselbaum2007}
\bibinfo{author}{\bibfnamefont{A.}~\bibnamefont{Weichselbaum}}
  \bibnamefont{and} \bibinfo{author}{\bibfnamefont{J.}~\bibnamefont{von
  Delft}}, \bibinfo{journal}{Phys.~Rev.~Lett.} \textbf{\bibinfo{volume}{99}},
  \bibinfo{pages}{076402} (\bibinfo{year}{2007}).

\bibitem[{\citenamefont{Peters and Pruschke}(2007)}]{peters2007}
\bibinfo{author}{\bibfnamefont{R.}~\bibnamefont{Peters}} \bibnamefont{and}
  \bibinfo{author}{\bibfnamefont{T.}~\bibnamefont{Pruschke}},
  \bibinfo{journal}{Phys.~Rev.~B} \textbf{\bibinfo{volume}{76}},
  \bibinfo{pages}{245101} (\bibinfo{year}{2007}).

\bibitem[{\citenamefont{Otsuki et~al.}(2009)\citenamefont{Otsuki, Kusunose, and
  Kuramoto}}]{otsuki2009}
\bibinfo{author}{\bibfnamefont{J.}~\bibnamefont{Otsuki}},
  \bibinfo{author}{\bibfnamefont{H.}~\bibnamefont{Kusunose}}, \bibnamefont{and}
  \bibinfo{author}{\bibfnamefont{Y.}~\bibnamefont{Kuramoto}},
  \bibinfo{journal}{J. Phys. Soc. Jpn.} \textbf{\bibinfo{volume}{78}},
  \bibinfo{pages}{034719} (\bibinfo{year}{2009}).

\bibitem[{\citenamefont{Efremov et~al.}(2008)\citenamefont{Efremov, Betouras,
  and Chubukov}}]{efremov2008}
\bibinfo{author}{\bibfnamefont{D.~V.} \bibnamefont{Efremov}},
  \bibinfo{author}{\bibfnamefont{J.~J.} \bibnamefont{Betouras}},
  \bibnamefont{and} \bibinfo{author}{\bibfnamefont{A.}~\bibnamefont{Chubukov}},
  \bibinfo{journal}{Phys.~Rev.~B} \textbf{\bibinfo{volume}{77}},
  \bibinfo{pages}{220401(R)} (\bibinfo{year}{2008}).

\bibitem[{\citenamefont{Tsunetsugu et~al.}(1997)\citenamefont{Tsunetsugu,
  Sigrist, and Ueda}}]{tsunetsugu1997}
\bibinfo{author}{\bibfnamefont{H.}~\bibnamefont{Tsunetsugu}},
  \bibinfo{author}{\bibfnamefont{M.}~\bibnamefont{Sigrist}}, \bibnamefont{and}
  \bibinfo{author}{\bibfnamefont{K.}~\bibnamefont{Ueda}},
  \bibinfo{journal}{Rev.~Mod.~Phys.} \textbf{\bibinfo{volume}{69}},
  \bibinfo{pages}{809} (\bibinfo{year}{1997}).

\bibitem[{\citenamefont{Smerat et~al.}(2009)\citenamefont{Smerat, Schollwöck,
  McCulloch, and Schoeller}}]{smerat2009}
\bibinfo{author}{\bibfnamefont{S.}~\bibnamefont{Smerat}},
  \bibinfo{author}{\bibfnamefont{U.}~\bibnamefont{Schollwöck}},
  \bibinfo{author}{\bibfnamefont{I.~P.} \bibnamefont{McCulloch}},
  \bibnamefont{and}
  \bibinfo{author}{\bibfnamefont{H.}~\bibnamefont{Schoeller}},
  \bibinfo{journal}{Phys.~Rev.~B} \textbf{\bibinfo{volume}{79}},
  \bibinfo{pages}{235107} (\bibinfo{year}{2009}).

\end{thebibliography}

\end{document}